\documentstyle[amsfonts,aps]{revtex}

\twocolumn
\begin{document}

%\draft   

\title{Comment on ``Atomic jumps in quasiperiodic Al$_{72.6}$Ni$_{10.5}$Co$_{16.9}$
and related crystalline material''} 
\author{Gerrit Coddens}

\address{Laboratoire des Solides Irradi\'es, Ecole Polytechnique,
F-91128-Palaiseau CEDEX, France}   % Declares the author's name.
\date{2nd December2002}   % Deleting this command produces today's date.

\maketitle

\widetext 
\begin{abstract}
We disagree with a number of statements by Dolin{\v{s}}ek et al. about 
the specificity of phason dynamics in quasicrystals (QCs).\\
\end{abstract}

\narrowtext 

We are surprised by the lack of foundation and clarity
of some formulations in a recent paper by Dolin{\v{s}}ek et al.\cite{Dolinsek}
about phason dynamics in QC.
It must be clear that there is no ground for an allegation that
the dynamical signal observed in references\cite{phasons}
would be due to vacancy diffusion rather than phason hopping.
In fact, the $Q$-dependence of the neutron-scattering signal
indicates that the atomic motion remains confined in space,
while its temperature dependence is unusual and
not typical of vacancy diffusion. Furthermore,
the diffusion constants
that one experimentally observes are much lower than
the ones one should reasonably expect on the basis
of the observed hopping rates if this hopping 
were  due to vacancy diffusion.
Indeed, the hopping is exceptionally fast,
while the observed diffusion constants just take  values
that could be qualified as standard for 
metallic compounds. The authors are stating this themselves,
such that they do not seem to question all these matters.
It is therefore puzzling
what the  statement about a
remarkable similarity
between the activation energy of the hopping process
and the enthalpy for the formation of a vacancy in pure Al
is supposed to imply. What is the aim of this statement
about what probably is merely a numerical coincidence?
We must mention  that the numbers 
quoted from our work  are {\em not} the {\em activation} 
energies of the hopping process, but {\em assistance} 
energies.
Moreover, the value of 0.6 eV quoted for AlCuFe is related with Cu 
rather than with Al hopping, and
the assistance energies we reported for AlPdMn are
not remarkably close to 0.6 eV. 
On the same footing it is not clear  what the statement
that high temperature phason hopping
would ``interfere'' with other processes is supposed to imply.

The authors claim that it has been shown that
d-AlCoNi contains a large amount of ``vacant sites''.
The``vacant sites'' in question are just phason sites,\cite{deca}
such that this claim contradicts the earlier introductory statement
of the authors that no empty lattice site is involved in 
the concept of a phason flip.
That we are dealing here with phason sites rather than with vacancies
is not an issue and leaves no space for any confusion of the
kind that would seem to  emanate from the presentation of the authors.
The confusion is produced by the undifferentiated terminology
``vacant site'' which the authors use for both phason sites and vacancies.
% This transpires already from the fact that the authors cite in this context
% the ``pipe mechanism'' of G\" ahler and Roth\cite{Gaehler}
% which is a phason-mediated (as opposed to
% vacancy-mediated) diffusion mechanism (of which G\" ahler and Roth
% also discuss a vacancy-{\em assisted} variant)\\

Other  poorly justified statements have been made 
by these authors to promote
the idea that QCs would contain a large amount
of vacancies.\cite{Dubois1,comment1} 
This shows that\\

~\\

\noindent this statement has not been
deduced by unbiased logical deduction from scientific observation,
but that it is a preconceived postulate the 
authors try to validate.
This postulate has a pivotal function in the argument
of the authors. It must  serve
to lend credibility to an analogy they want to impose between
B2-based crystals (which might contain up to 12 percent structural
vacancies) and QCs (where nothing of 
that order of magnitude has ever 
been established).\cite{footnote}
This then should permit to incorporate phason dynamics
into a much broader class of trite hopping phenomena.

The authors argue that in contrast with high-temperature
data, the low-temperature data would not be subject to ``interference''. 
This cannot conceal the lack of conclusive 
prove for the attribution of the 
low-temperature NMR signal to phason dynamics, as
 reflected in the caution of the 
statement that the data are {\em compatible} with phason dynamics.
NMR data do not provide much information  about
characteristic distances as they do not yield
Q-dependent information. The time scales accessed are quite
remote from the time scales that can be accessed by other techniques
such that cross-checking or use of complementary information to
validate the claims is not possible.
This is unfortunate as the existence of such slow low-temperature
dynamics is certainly intriguing.  
The only thing we really know with certainty is 
that there is some slow (local) relaxation 
with a low activation energy. E.g. the idea of a small 
local shift of the Al atom
in response to a slowly fluctuating or diffusing 
strain field  that includes its
environment is equally compatible with their data.
One can think of other phenomena that could be compatible
with the data. Without wanting to reinterpret the NMR data
accordingly,  we may mention that one type or another of
slow dynamics remaining unfrozen at low temperatures 
is observed in many systems,
including QCs, e.g. in the form of tunneling states.\cite{Bert}
Above a few degrees K the
coherent tunelling may cross over to a thermally activated process.
The low temperature region is thus not as exempt of the possibility
of ``interfering'' dynamics as
the authors suggest.
It has still not been proved experimentally
that tunneling states in QCs are not phasons. 
This illustrates the difficulty of making assignments.
In glasses tunneling states are conceived as 
small simultaneous shifts in the 
positions of groups of atoms bringing about a transition
between slightly degenerate configurations.
The precise detailed geometrical picture
of the motions involved in such processes remains unknown.
This kind of lack of information  is
a recurrent theme in slow dynamics and
due to the  physical limitations of the available 
experimental techniques. From all this we may retain a 
neglected possibility of explanations in terms
of small, non-phasonic atomic shifts,
whose amplitudes explore a continuum rather than a discrete set.

The suggested uniqueness of interpretation of
the NMR data is also contradicted by other  NMR data  
of the authors.
In fact, in reference \cite{NMR} they observed a temperature
behaviour of an NMR signal that was not compatible
with their interpretation of it in terms of phason dynamics. 
The AlPdMn phase can be magnetic, a fact that 
offers seeds for an alternative
explanation.\cite{private}
In a subsequent paper they stated\cite{coverup} 
that this scenario could not be proved, 
and concluded that
the origin of the signal was not understood, 
but probably due to
unusual magnetic properties of Mn atoms.
They also stated that the data from AlPdRe are of a
different origin than those from AlPdMn, despite possible 
similarities in the time scales.
This illustrates 
that other phenomena that are not frozen can exist 
at low temperatures and 
wrongfoot the interpretation of the data.
It is not because two signals look similar that they cover
similar physics: What is more similar to a 
relaxation time with
an activation energy than another relaxation 
time with an activation energy?
But in many experimental techniques, relaxation 
times and activation energies are  
all the hard data resume to.
Now the authors have established the existence of 
similar signals  in B2-based AlNiCu, 
which 
despite all possible claims definitely does not support
phason dynamics. If following them we discard 
interference from vacancy motion, we are faced with the problem
of elucidating the origin of an NMR signal in the B2-based phase
that is not due to  phasons nor to vacancy hopping. The fact 
that such unclearly identified atomic dynamics
 can give rise 
to a signal that is similar 
to the one observed in QCs is 
(in view of the unappropriateness of the comparison)
not a proof that the assignment
of the authors is wrong, but does reveal that its 
uniqueness  is not sufficiently 
substantiated.
We may note that the theory of Jari\'c and 
Nelson  for diffuse scattering\cite{Jaric}
is based on the ansatz that  phason dynamics is frozen.
Hence the assignment of the authors also summons for a 
rethinking of matters that have been directly linked to the  validity 
of  the random tiling model. 

In summary, the main conclusion 
of the authors that phason dynamics
is not QC-specific cannot be reached on the
basis of the data presented. But even if 
such a conclusion 
could be reached, nobody would
understand QC-specifity in the 
singular, restrictive sense 
 the authors want to give to it.
 Their concern of uniqueness is 
 much less in order than
 the one about the assignment of the NMR signal.
In fact, phasons correspond to 
atomic jumps in double-well potentials,
whose minima are separated by a distance 
shorter than the interatomic distances.
Nobody claims that such double-well 
potentials giving rise to atomic jumps 
would be a rarity in solid state physics. 
The example of the hydrogen bond has been well known 
for a long time. One of the original features of
QCs is that the presence of these short-distance double 
wells is an integrated part  of the
quasilattice. Conceptually, a simple B2-based lattice 
with vacancies does not 
imply the existence of short-distance 
double wells. 
It would thus appear that examples of
signals from  B2-based phases do not 
capture the essence
of the non-uniqueness of the dynamics of 
such double-well potentials.
As already stated, there is more 
to the similarity of the physics
than a similarity of relaxation times.

\end{document}